\documentclass{bmcart}

\usepackage{amsthm,amsmath}
\usepackage[utf8]{inputenc} 

\usepackage{booktabs}
\usepackage{multirow}
\usepackage{array, makecell} 

\usepackage{caption}
\usepackage{subcaption}

\usepackage{graphicx}

\startlocaldefs
\endlocaldefs

\begin{document}

\begin{frontmatter}

\begin{fmbox}
\dochead{Research}


\title{Visualizing adverse events  in clinical trials using correspondence analysis with R-package visae}


\author[addressref={aff1}, 
  corref={aff1},            
  email={marcio.diniz@cshs.org}   
]{\inits{MAD}\fnm{M{\'a}rcio A.} \snm{Diniz}}
\author[
  addressref={aff1},
  email={gillian.gresham@cshs.org}
]{\inits{GG}\fnm{Gillian} \snm{Gresham}}
\author[
  addressref={aff1},
  email={sungjim.kim@cshs.org}
]{\inits{SK}\fnm{Sungjin} \snm{Kim}}
\author[
  addressref={aff1},
  email={michael.luu@cshs.org}
]{\inits{ML}\fnm{Michael} \snm{Luu}}
\author[
  addressref={aff2},
  email={norahh@med.umich.edu}
]{\inits{NLH}\fnm{N. Lynn} \snm{Henry}}
\author[
  addressref={aff1},
  email={mourad.tighiouart@cshs.org}
]{\inits{MT}\fnm{Mourad} \snm{Tighiouart}}
\author[
  addressref={aff3},
  email={yothersg@nrgoncology.org}
]{\inits{GY}\fnm{Greg} \snm{Yothers}}
\author[
  addressref={aff4},
  email={pganz@mednet.ucla.edu}
]{\inits{PG}\fnm{Patricia A.} \snm{Ganz}}
\author[
  addressref={aff1},
  email={andre.rogatko@cshs.org}
]{\inits{AR}\fnm{Andr{\'e}} \snm{Rogatko}}


\address[id=aff1]{
  \orgdiv{Samuel Oschin Compreheensive Cancer Center},             
  \orgname{Cedars-Sinai Medical Center},          
  \city{Los Angeles, CA},                              
  \cny{US}                                    
}

\address[id=aff2]{
  \orgdiv{Rogel Cancer Center},             
  \orgname{University of Michigan},          
  \city{Ann Arbor, MI},                              
  \cny{US}                                    
}

\address[id=aff3]{
  \orgdiv{Graduate School of Public Health},             
  \orgname{University of Pittsburgh and NRG Oncology},          
  \city{Pittsburgh, PA},                              
  \cny{US}                                    
}

\address[id=aff4]{
  \orgdiv{Fielding School of Public Health},             
  \orgname{UCLA Fielding School of Public Health},          
  \city{Los Angeles, CA},                              
  \cny{US}                                    
}



\end{fmbox}


\begin{abstractbox}

\begin{abstract} 

\parttitle{Background} 
Graphical displays and data visualization are essential components of statistical analysis that can lead to improved understanding of clinical trial adverse event (AE) data. Correspondence analysis (CA) has been introduced decades ago as a multivariate technique that can communicate AE contingency tables using two-dimensional plots, while quantifying the loss of information as other dimension reduction techniques such as principal components and factor analysis. 

\parttitle{Methods} 
We propose the application of stacked CA using contribution biplots as a tool to explore differences in AE data among treatments in clinical trials. We defined five levels of refinement for the analysis based on data derived from the Common Terminology Criteria for Adverse Events (CTCAE) grades, domains, terms and their combinations. In addition, we developed a Shiny app built in an R-package, visae,publicly available on Comprehensive R Archive Network (CRAN), to interactively investigate CA configurations based on the contribution to the explained variance and relative frequency of AEs. Data from two randomized controlled trials (RCT) were used to illustrate the proposed methods: NSABP R-04, a neoadjuvant rectal 2x2 factorial trial comparing radiation therapy with either capecitabine (Cape) or 5-fluorouracil (5-FU) alone with or without oxaliplatin (Oxa), and NSABP B-35, a double-blind RCT comparing tamoxifen to anastrozole in postmenopausal women with hormone-positive ductal carcinoma in situ.
\parttitle{Results} 
In the R04 trial (n=1308), CA biplots displayed the discrepancies between single agent treatments and their combinations with Oxa at all levels of AE classes, such that these discrepancies were responsible for the largest portion of the explained variability among treatments. In addition, an interaction effect when adding Oxa to Cape/5-FU was identified when the distance between Cape+Oxa and 5-FU+Oxa was observed to be larger than the distance between 5-FU and Cape, with Cape+Oxa and 5-FU+Oxa in different quadrants of the CA biplots. In the B35 trial (n=3009), CA biplots showed different patterns for non-adherent Anastrozole and Tamoxifen compared with their adherent counterparts.
\parttitle{Conclusion} 
CA with contribution biplot is an effective tool that can be used to summarize AE data in a two-dimensional display while minimizing the loss of information and interpretation. 
\end{abstract}




\end{abstractbox}
%

\end{frontmatter}



\section*{Background}

The understanding of adverse events is paramount in the assessment of therapies in clinical trials. In the endeavor to support investigators in the challenging task of identifying and documenting toxicities, the National Cancer Institute has maintained, since 1983, an empirical lexicon of AE terms that are commonly encountered in oncology: the CTCAE \cite{ctcaev4}, which has been broadly adopted over the last decades. The criteria classifies AE into 26 domain organ classes and severity grades, such that grade 1 corresponds to a mild or asymptomatic symptom and grade 5 indicates death. 

Although such comprehensive criteria has allowed investigators to collect a large amount of AE clinical trial data, the abundance of information has often been ignored in the analysis of clinical trials. Analyzing AE data is a complex task because each patient could experience more than one AE term from different organ domains and different grades of the same AE term during several cycles of treatment resulting in a high-dimensional toxicity profile. Investigators usually present lengthy and overwhelming AE tables, or partial toxicity profiles by treatment after summarizing AE chosen based either on  a frequency threshold, severity of AEs or  relatedness to treatments \cite{phillips2019analysis} into their maximum grade. However, the use of maximum grade leads to loss of information and has been largely criticized and other alternative approaches have been discussed in the literature that summarizes toxicity profiles into a more comprehensive score \cite{lee2011, thanarajasingam2016, gresham2020}. 

The CONSORT extension for reporting harm outcomes \cite{ioannidis2004better} describes the importance of graphical displays for summarizing AE data. Several graphical summary approaches have been developed in the literature since then: Amit et al. \cite{amit2008} considered dot-plots for the percentage of AE terms' occurrence by treatment ordered by their relative risks; Zink et al. \cite{zink2013} proposed volcano plots with bubble size proportional to the frequency of AE domain or terms; Thanarajasingam et al. \cite{thanarajasingam2016} recommended profile plots to illustrate the average toxicity as function of cycle for a given AE term; Karpefors and Weatherall \cite{karpefors2018} suggested tendril plots to represent the occurrence of a given AE term over time, and Gresham et al. \cite{gresham2020} proposed stacked barplots for the AE frequency as function of the number of toxicities per patient and grade toxicity. However, the majority of these approaches cannot be applied to more than two treatments.

Surprisingly, none of the aforementioned approaches have used any traditional statistical high dimension reduction technique such as CA, which is a multivariate technique with the purpose to communicate contingency tables using two-dimensional graphical displays, while quantifying the loss of information. Initial applications of CA and its variants (stacked, multiple, detrended) were broadly discussed in Greenacre \cite{greenacre1992correspondence}, with specific applications in epidemiology \cite{sourial2010correspondence, hirsch2011multivariate, befus2017correspondence} and bioinformatics \cite{fellenberg2001correspondence, busold2005integration, horita2020time}. 

In this article, we propose the use of stacked CA as a visualization tool for AE data to unravel differences in treatment profiles when comparing their AEs as a complementary tool to toxicity scores \cite{gresham2020, le2020adverse}. An R-package was developed to make our approach available interactively. We illustrate the use of CA to identify different toxicity profiles among treatments in two clinical trials R04 \cite{russell2015, allegra2015} and B35 \cite{cella2008, land2011, margolese2016, ganz2016, land2016}. Moreover, we apply CA using contribution biplots \cite{greenacre2013} that are not widely disseminated yet, even though they address long-standing interpretation issues of CA such as outlier points with low  contribution to the variance. 


\section*{Methods} \label{sec:ca}

\subsection*{Correspondence Analysis}

The seminal ideas of CA was proposed by Herman Otto Hartley (Hirschfeld) \cite{hirschfeld1935}, later developed by Jean-Paul Benzécri \cite{benzecri1973} and disseminated by Michael Greenacre \cite{greenacre1984, greenacre2017}. We will briefly review the main concepts of CA. A detailed mathematical is available as an Additional file. 

The goal of CA is to graphically represent contingency tables.  Following Greenacre \cite{greenacre2017}, we will apply CA on stacked tables such as Table \ref{tab:contingency_table03}. 

\begin{table}[!ht]
\centering
\begin{tabular}{ccccc}
\toprule
\multirow{2}{*}{AE class} & \multicolumn{3}{c}{Treatment (T)} & 
\multirow{2}{*}{Total per AE}\\
                     & 1 & $\ldots$ &  J & \\
\midrule
1                     & $\pi_{11}$ & $\ldots$ & $\pi_{1J}$ & $\sum_{j = 1}^J\pi_{1j}$\\
\hline
$1^c$                    & $1 - \pi_{11}$ & $\ldots$ & $1 - \pi_{1J}$ & $J - \sum_{j = 1}^J\pi_{1j}$\\
\hline
$\vdots$             & $\vdots$ & $\vdots$ & $\vdots$ & $\vdots$ \\
\hline
I                     & $\pi_{I1}$ & $\ldots$ & $\pi_{IJ}$ & $\sum_{j = 1}^J\pi_{Ij}$\\
\hline
$I^c$                    & $1 - \pi_{I1}$ & $\ldots$ & $1 - \pi_{IJ}$ & $J - \sum_{j = 1}^J\pi_{Ij}$\\
\hline
Total per treatment      & I & $\ldots$ & I & IJ\\
\bottomrule
\end{tabular}
\caption{Contingency table $2I \times J$ with row and column marginals}
\label{tab:contingency_table03}
\end{table}

\noindent  where $\pi_{ij}$ is the  relative frequency of $AE_i$ class for treatment $T_j$ such that AE classes can be based on three levels of data aggregation: (a) AE grades, (b) AE domains, (c) AE terms and their combinations. While toxicity profiles can be presented and compared based on tables when AE classes are only defined by AE grades, it is a much more complex task when AE classes are defined by AE domains or AE terms, and their combinations with AE grades. There are 26 domains and 790 AE terms in CTCAE v4, which can generate 130 AE classes when AE domains are combined with AE grades and 3950 AE classes when AE terms are combined with AE grades.

The interpretation of Table \ref{tab:contingency_table03} is asymmetric: we are interested in studying the differences in toxicity profiles among treatments that lie in a high-dimensional space, which will be denoted as toxicity space. Visualizing the toxicity profiles in the toxicity space can give us insight regarding the association between treatments and AE classes. Nonetheless, it is not always feasible to display toxicity profiles when the number of dimensions is greater than three, i.e., four treatment arms $(J \geq 4)$ or four AE classes ($I \geq 4$). Moreover, distances between toxicity profiles of treatments are not simple to be evaluated even in a three dimensional space. 

In this context, CA seeks the two-dimensional display that minimizes the loss of information when reducing the dimension of the toxicity space. Information is measured through variability, denoted as total inertia in CA, among toxicity profiles of treatments. Asymmetric contribution biplots \cite{greenacre1993} are two-dimensional diplaying showing the projection of the two dimensions with  highest variability in the toxicity space. The first dimension of the biplot represents the direction with highest variability of the toxicity space and the second dimension corresponds to the direction with the second highest inertia. Adding up the inertia of the remaining dimensions allows us to quantify the loss of information. Therefore, we can evaluate whether the two-dimensional representation of the toxicity space is adequate.

The inertia associated with each dimension can also be broken down based on the contributions of each AE class. Dimensions can be interpreted based on AE classes with high contributions. AE classes with high contributions to a dimension arex` identified based on their distance from the origin in the same direction of that given dimension. Then, a treatment with high frequency of a given AE class will have high values in the same dimension and direction of that AE class.  

In this way, we are able to compare toxicity profiles of treatments as following:
\begin{enumerate}
\item Interpret each dimension based on the position of AE classes dots: AE classes further away from origin (0, 0) in a given direction indicates a high contribution to explain the variability in that dimension;
\item Identify level of similarity among toxicity profiles of treatments based on how close their toxicity profiles are from the origin (0, 0), which represents a hypothetical average toxicity profile; 
\item Compare toxicity profiles of treatments based on their position on each dimension. 
\end{enumerate}
Notice that distances between treatment profiles and AE classes are not meaningful because they are different space. Figures S1 and S2 based on toy data illustrate these steps.

\subsection*{R-package} \label{sec:rpackage}

We developed the R-package \emph{visae}, an acronym for visualizing AE, aiming to provide statistical software to quickly deploy Shiny applications making our visual approach interactively available for AE reporting. Currently, there are two R-packages specific for CA: \emph{ca} \cite{ca2007} and \emph{FactoMineR} \cite{factominer2008}. The R-package \emph{visae} is built based on \emph{ca}. Although both R-packages ca and \emph{FactoMineR} provides CA biplots, the R-package visae makes available the pre-processing of AE data to construct tables such as Table \ref{tab:contingency_table03} and interactive Shiny application to explore CA configurations.

Interactive applications allow statisticians and non-statisticians to easily collaborate to investigate several configurations for CA, and select the ones that are more informative to them.  Therefore, the R-package \emph{visae} provides a general framework for CA allowing statisticians easily interact with their collaborators.

The function \textit{run\_ca} has seven arguments, with four of them required to execute the Shiny application:
\begin{itemize}
    \item \textit{data}: a data.frame or tibble object in a long format;
    \item \textit{group}: unquoted variable name in the data that corresponds to the group variable;
    \item \textit{id}: unquoted variable name in the data that corresponds to the patient identification variable;
    \item \textit{ae\_grade}: unquoted variable name in the data that corresponds to AE grade class;
\end{itemize}

While the other three inputs can be used in any combination,
\begin{itemize}
    \item \textit{ae\_domain}: unquoted variable name in the data that corresponds to AE domain class;
    \item \textit{ae\_term}: unquoted variable name in the data that corresponds to AE term class;
    \item \textit{ae\_cycle}: unquoted variable name in the data that corresponds to AE cycle.
\end{itemize}

For example, an R user can open the Shiny application as below:
\begin{verbatim}
library(visae)
library(magrittr)

patient_id <- 1:100
group <- c(rep("A", 50), rep("B", 50))
grade <- sample(1:5, size = 100, replace = TRUE)
domain <- sample(c("C", "D"), size = 100, replace = TRUE)
term <- sample(c("E", "F", "G", "H"), size = 100, replace = TRUE)

data <- data.frame(patient_id, group, grade, domain, term)
head(data, n = 6)

  patient_id group grade domain term
1          1     A     3      C    G
2          2     A     3      C    E
3          3     A     2      D    G
4          4     A     2      D    E
5          5     A     5      D    F
6          6     A     3      C    H

data %>% run_ca(group = trt, 
                id = patient_id,
                ae_grade = grade, 
                ae_domain = domain,
                ae_term = term)
                
\end{verbatim}
All the contribution biplots and relative frequency tables presented in the next sections were generated using our Shiny application. 


\subsection*{Data sets} \label{sec:data}
Data from two randomized clinical trials from the National Surgical Adjuvant Breast and Bowel Project (NSABP) were used as case examples for this analysis: 

\subsubsection*{R04}

NSABP R04 was a Phase III randomized 2x2 factorial trial comparing neoadjuvant radiation therapy (RT) in combination with either Cape or 5-FU with or without Oxa in patients with rectal cancer (NCT00058474)\cite{allegra2015}. AE data (CTCAE version 4.0) were collected at a single time point and included a list of 50 AEs of special interest that were selected a priori and evaluated after chemoradiation treatment within 2 weeks of surgery. Additional details of the trial are reported elsewhere \cite{allegra2015}.

\subsubsection*{B35}

NSABP B35 was a Phase III double-blind, randomized, placebo-controlled trial comparing daily oral tamoxifen with oral anastrozole for 5 years in postmenopausal women with hormone receptor-positive ductal carcinoma in situ treated with lumpectomy and radiation therapy (NCT00053898)\cite{forbes2016anastrozole}.  Adverse events were assessed every 6 months during therapy and 6 months  after the last dose of therapy using a list of predefined AEs (e.g., depression, thromboembolic events, GI disturbance, hot flashes, joint pain, vaginal dryness), graded per Common Toxicity Criteria (CTC) v2.0. Non-adherent patients were defined as patients that stopped treatment early before 5 years for reasons other than disease progression or death.

\section*{Results} \label{sec:application}

We illustrate the main concepts of correspondence analysis when used to represent AE data comparing treatments in the R04 trial and discuss the interaction between treatment and adherence using data from the B35 trial. Analyses at five levels of refinement are presented, but only contingency tables for AE classes defined based on grades are shown. For all other AE classes, contingency tables are presented as Additional File. 

\subsection*{R04}

We performed CA considering the different AE class definitions discussed previously with four treatments: 5-FU, Cape, 5-FU+Oxa and Cape+Oxa. Table \ref{tab:r04_ae_classes} shows the total number of AE classes for each definition with the percentage of explained inertia for each of the three dimensions. When we define AE classes solely based on AE grades, Table \ref{tab:contingency_table03} will have 10 ($2 \times I$) rows with 5 AE grades (each one adds its complementary), such that  a two-dimensional display describes 98.23\% of the variability among treatments with a 1.77\% loss of information when dimension 3 of the toxicity space is ignored. The loss of information when representing the toxicity space in a two-dimensional display increases as the level of complexity for AE classes increases. For all AE class definitions, the loss of information is no more than 21\% making the two-dimensional display an acceptable representations of the toxicity space. 

\begin{table}[!ht]
\centering
\begin{tabular}{cccccc}
\toprule
AE Class & $\#$ AE classes & Dim 1 & Dim 2 & Dim 3\\
\hline
Grade & 5 & 87.77 & 10.46 & 1.77\\
\hline
Domain & 21 &  84.65 & 10.97 & 4.38\\
\hline
Domain + Grade & 61 & 66.47 & 22.94 & 10.59\\
\hline
Term & 209 & 52.29 & 28.03 & 19.68\\
\hline
Term + Grade & 313 & 48.92 & 30.26 & 20.82\\
\bottomrule
\end{tabular}
\caption{AE classes and their decomposition of total inertia into 3 dimensions for R04 trial}
\label{tab:r04_ae_classes}
\end{table}

Initially, we assume AE classes based on AE grades as showed in Table \ref{tab:r04_ae_grade_percentage}. In Figure \ref{fig:r04_biplot}.a, main differences are observed in dimension 1 such that  discrepancies among treatments are small because their treatment profiles are posed close to each other and they are near to the origin, which represents the average treatment. In dimension 1, ignoring grade 1 AEs that were under-reported, treatments can be ordered based on their relative positions indicating that 5-FU is associated with the lowest frequencies for all AE grades, while Cape+Oxa and 5-FU+Oxa present higher frequency of grade 2, 3, 4 and 5 AEs than their corresponding single agents. Moreover, Cape is associated with higher frequency of grade 5 AEs than 5-FU, Cape+Oxa is associated with higher frequency of grades 1, 4 and 5 AEs, and 5-FU+Oxa with higher frequency of grade 2 AEs than other treatments.  

\begin{table}[!ht]
\centering
\begin{tabular}{cccccc}
\toprule
\multirow{2}{*}{AE Class} & \multicolumn{4}{c}{Treatment} & \multirow{2}{*}{Average}\\
& 5-FU & 5-FU + Oxa & Cape & Cape + Oxa & \\
\hline
G1 & 1.22 & 2.75 &	1.23 &	3.96 &	2.29\\
\hline
G2 & 60.67 & 74.01 & 63.08	& 70.73 & 67.12\\
\hline
G3 & 25.31 & 38.53	& 27.39 & 39.94 & 32.79\\
\hline
G4 & 0.61 & 3.06 & 2.15 & 4.27 & 2.52 \\
\hline
G5 & 0.31 & 0.31 & 1.23 & 1.52 & 0.84\\
\bottomrule
\end{tabular}
\caption{ Percentage (\%) of AE grades by treatment in R04 trial}
\label{tab:r04_ae_grade_percentage}
\end{table}

While Table \ref{tab:r04_ae_grade_percentage} is small enough to be understood without a CA biplot, we are interested in more refined AE classes. We define AE classes based on domains as showed in Figure \ref{fig:r04_biplot}.b, with main differences among treatments in dimension 1. Single agents 5-FU and Cape are not very different between them, but both of them differ from their combinations with Oxa. Treatment combinations are associated with AEs in the domains Immune, Nervous, General, Metabolism, Gastrointestinal and Investigations; such that Cape+Oxa and 5-Fu+Oxa are associated in a larger extent with AEs in the domain Investigations and Gastrointestinal, respectively. In addition, Cape+Oxa is associated with the domain Vascular and 5-FU+Oxa is associated with domains Infections and Hepatobiliary.

In Figure \ref{fig:r04_biplot}.c, AE classes are defined based on the combination between domains and grades, with differences in both dimensions. Similar interpretation as Figure \ref{fig:r04_biplot}.b can be outlined, except that the domains Metabolism and General are posed into different quadrants when broken down by grades: Metabolism:G2 and General:G2 are associated with 5-FU+Oxa, while Metabolism:G3 and General:G3 with Cape+Oxa. Moreover, it is possible to observe clustering of domains: (i) domains in the top left quadrant have higher frequency among patients that received 5-FU+Oxa; (ii) domains in the bottom left quadrant have higher frequency for Cape+Oxa; (iii) domain Nervous:G2 is associated with both treatments, and Injury:G2 with their single agents counterparts.

Next, we show AE terms in Figure \ref{fig:r04_biplot}.d with differences in both dimensions. As in the previous analyses, differences between Cape and 5-FU are small such that both are associated with higher frequency of Dermatitis RT when compared to their combinations with Oxa. Treatment combinations are both associated with several of AE terms including Peripheral sensory neuropathy, Diarrhea, Dehydration, Nausea, Vomiting, and Fatigue; such that 5-Fu+Oxa is associated in a larger extent with Diarrhea, Vomiting and Nausea. Cape+Oxa is also associated with Hand-foot syndrome.
In particular, anal pain and abdominal pain were not identified in CA configurations that were not overpopulated by AEs, indicating that their contribution to treatment differences is low in dimension 1 and 2 even though they have high average frequency of 19.42\% and 7.49\%, respectively. The highest contribution of anal pain is 1.22\% in the third dimension. 

Finally, we broke down the AE terms by adding their grades. In Figure \ref{fig:r04_biplot}.e, AE classes are defined based on terms and grades. Differences among treatments are found in both dimensions. When comparing Figure \ref{fig:r04_biplot}.e to Figure \ref{fig:r04_biplot}.d, we highlight the term Fatigue that was divided into Fatigue:G2 and G3 associated with 5-FU+Oxa and Cape+Oxa, respectively; the term Nausea was also divided into Nausea:G2 and G3 associated with 5-FU+Oxa and Cape+Oxa, respectively. Furthermore, the AE term Diarrhea is divided in Diarrhea:G3 associated with both 5-FU+Oxa dn Cape+Oxa, and Diarrhea:G2 associated with 5-Fu+Oxa.

Furthermore, the distance between 5-FU and Cape is smaller than the distance between 5-FU+Oxa and Cape+Oxa in all CA configurations, which can be interpreted as an interaction effect of Oxa. 

\subsection*{B-35}
  
We performed CA considering the different AE class definitions discussed in the previous section comparing four groups: adherent Anastrozole and Tamoxifen, and their non-adherent counterparts  based on AEs reported at cycle 1. Table \ref{tab:b35_ae_classes} shows the total number of AE classes for each definition with the percentage of explained inertia for each of the three dimensions.  For all AE class definitions, the loss of information is at most 11\% for the highest complexity level of the toxicity space.  

\begin{table}[!ht]
\centering
\begin{tabular}{cccccc}
\toprule
AE Class & $\#$ AE classes &  Dim 1 & Dim 2 & Dim 3\\
\hline
Grade & 3 & 92.57 & 6.98 & 0.45\\
\hline
Domain & 20 &  62.35 & 34.46 & 3.19\\
\hline
Domain + Grade & 60 & 55.73 & 39.80 & 4.47\\
\hline
Term & 214 &  47.20 & 43.01 & 9.79\\
\hline
Term + Grade & 384 & 48.18 & 41.52 & 10.30\\
\bottomrule
\end{tabular}
\caption{AE classes and their decomposition of total inertia into 3 dimensions for B35 trial}
\label{tab:b35_ae_classes}
\end{table}

As previously, we assume AE classes based on AE grades as showed in Table \ref{tab:b35_ae_grade_percentage}. Because of small size of contingency table, the first dimension is enough to understand the differences among treatments in Figure \ref{fig:b35_biplot}.a.  Adherent Tamoxifen and Anastrozole groups are quite similar to each other and are close to the average treatment. Both non-adherent groups are on the left quadrants indicating higher frequency of AE grades 2, 3 and 4. Nonetheless, non-adherent Tamoxifen and Anastrozole are in different quadrants showing that their discrepancies from the average treatment are grade specific: non-adherent Tamoxifen is more associated with grade 4 AEs, while non-adherent Anastrozole is more associated with grade 2 and 3 AEs. Table \ref{tab:b35_ae_grade_percentage} leads to similar conclusions.

\begin{table}[!ht]
\centering
\begin{tabular}{cccccc}
\toprule
\multirow{2}{*}{AE class} & \multicolumn{2}{c}{Adherent} & \multicolumn{2}{c}{Non-adherent}& \multirow{2}{*}{Average}\\
& Anastrozole & Tamoxifen & Anastrozole & Tamoxifen & \\
\hline
G2 & 38.22 & 40.69	& 49.66 & 52.23 & 45.20 \\
\hline
G3 & 2.22 & 3.91 & 10.61 & 11.94 & 7.63\\
\hline
G4 & 0.188 & 0.279 & 0.903 & 3.279 & 1.162\\
\bottomrule
\end{tabular}
\caption{Percentage (\%) of AE grades at cycle 1 by treatment in B35 trial}
\label{tab:b35_ae_grade_percentage}
\end{table}

In Figure \ref{fig:b35_biplot}.b, AE classes are defined based on domains, such that  differences between adherent and non-adherent groups are showed in the first dimension and differences between non-adherent Anastrozole and non-adherent Tamoxifen in the second dimension. Non-adherent groups are associated with higher frequency of AEs in the domains Pain, Neurology, Constitutional Symptoms, Psychiatric, Cardiovascular, Gastrointestinal, Allergy/Immunology. In particular, non-adherent Anastrozole is more associated with the domains Psychiatric and Pain, while non-adherent Tamoxifen is associated with the domains Gastrointestinal, Allergy/Immunology and Cardiovascular.


Figure \ref{fig:b35_biplot}.c combines AE grades to the AE domains as AE classes. Differences are found in both dimensions with similar interpretation from Figure \ref{fig:b35_biplot}.b. In particular, we highlighted the domain Pain from Figure \ref{fig:b35_biplot}.b that was broken down into Pain:G3 associated with both non-Aaherent Tamoxifen and  non-adherent Anastrozole, while Pain:G2 is associated only with non-adherent Anastrozole..

In Figure \ref{fig:b35_biplot}.d, AE classes are defined based on terms. Differences are found in both dimensions with similar interpretation from Figure \ref{fig:b35_biplot}.b. The AE terms can be divided into three clusters: (i) Dizziness, Sweating, Edema, Constipation, Dyspnea, Hot flashes and Radiation dermatitis associated with non-adherent Tamoxifen; (ii) Arthralgia, Bone Pain, Headache, Myalgia, Varginal dryness are associated with non-adherent Anastrozole; (iii) Fatigue, Depression,  and Insomnia are associated with both non-adherent groups.

As the last step, Figure \ref{fig:b35_biplot}.e combines terms and grades. There is a change in the interpretation of the dimensions: discrepancies between non-adherent Anastrozole and Tamoxifen becomes more relevant to explain the variability among groups than the differences between adherent and their non-adherent counterparts, which yields the reverse interpretation of the dimensions from Figures \ref{fig:b35_biplot}.a-\ref{fig:b35_biplot}.d. Furthermore, clusters of AEs associated with non-adherent Anastrozole and Tamoxifen are observed similar to Figure \ref{fig:b35_biplot}.d.

\section*{Discussion} \label{sec:discussion}

We proposed stacked CA using contribution biplots as a tool to explore differences in AE data among treatments in clinical trials. We defined five levels of refinement for the analysis based on AE grades, domains, terms and their combinations. In addition, we developed a Shiny application built in an R-package to interactively investigate CA configurations based on the contribution to the explained variance and relative frequency of AEs, and we made it publicly available on CRAN. Phillips et al \cite{phillips2019analysis} have found that only 12\% among 184 clinical trials published in major medical journals between 2015 and 2016 showed graphical presentations. We expect to improve AE reporting through statistical graphical displays and easy-to-use software that can also be transformed into web applications as suggested by the Consolidated Standards of Reporting Trials (CONSORT) Harm extension \cite{ioannidis2004better}. Furthermore, the proposed analysis could also be applied to patient reported outcomes (PRO) such as PRO-CTCAE \cite{dueck2015validity}.

Morever, we illustrated the use of stacked CA for different goals in the R04 and B35 clinical trials. In our examples, toxicity spaces representing contingency tables at the highest level of refinement of AE classes - term and grade combination - have dimension three such that the loss of information when using CA biplots was 20.82\% in R04 and 10.30\% in B35, which are within the threshold of 30\% as proposed by some authors. In R04 trial, CA biplots displayed the differences in AE patterns between single agent treatments and their combinations with Oxa at all levels of AE classes, such that an interaction effect when adding Oxa to Cape/5-FU was identified. In the B35 trial, CA biplots showed the discrepancies between non-adherent and adherent Tamoxifen and Anastrozole. Different patterns for non-adherent Anastrozole and Tamoxifen were observed contrasting with their adherent counterparts. Interestingly, CA biplots identified expected differences in AE frequency between treatments (e.g., Arthralgia associating with non-adherent Anastrozole and Hot Flashes associated with non-adherent Tamoxifen), but also others that were less expected (e.g., Depression and fatigue associated with Anastrozole and Dizziness with Tamoxifen).

The main goal of CA in the context of AEs is to visualize associations between treatments and AEs while controlling the loss of information due the dimension reduction of the toxicity space. The loss of information is an increasing function of the level of refinement of AE classes. The several levels of AE classes could give researchers insights regarding the discrepancies among treatments. While CA with AE grades presents a more understandable biplot, it does not discriminate the toxicity profiles enough. On the other hand, CA biplots with AE terms and grades presents a lot of information that might make harder to draw conclusions at the same time allowing one to understand detailed differences among groups. A possible compromise is CA biplots with domains or domains and grades that indicate enough differences among treatment without being visually overwhelming. 

Ideally, CA biplots with AE classes defined based on AE grades and their combinations with terms/ domains would show AEs following the grade ordering. Nonetheless, such pattern is rarely observed even in CA biplots that are based solely on AE grades with large variability explained by dimension 1. The lack of ordering is expected, although not intuitive, because associations,  which are shown by CA biplots, between treatments and AE classes defined based on AE grades are often not ordered. At first glance, the understanding of CA biplots might be misleading, but we believe that annotated toy CA biplots as presented in this work will help researchers to interpret results. 

We have not discussed inference based on CA given its limited scope. Few authors have \cite{ringrose2012bootstrap, lombardo2012bootstrap, beh2015confidence} presented inferential procedures using bootstrap to provide confidence regions on CA biplots with poor performance when dealing with sparse matrices. Also, we did not study the association pattern within AE classes for a treatment and multiple AE of the same grade are not taken into account if a CA biplot is based on more than one treatment cycle. In future work, we plan to apply joint CA to visualize association pattern within AE classes by treatments and analysis of matched matrices to compare such patterns between treatments, respectively. 

\section*{Conclusion}

CA with contribution biplot is an effective tool that can be used to summarize AE data in a two-dimensional display while minimizing the loss of information and interpretation.  It is general enough to be applied to a variety of drugs classes and diseases. Instead of lengthy frequency tables presented as supplemental material of trial reports, CA biplots for AE classes defined based on either terms or the combination between terms and grades could be presented, so the data can be examined visually. Furthermore, CA could be used to help investigators to select AE classes to be summarized based on objective criteria given by their contributions to the explained variance and relative frequencies. In this way, AE reporting would be more consistent across studies. A drawback of such strategy is that it could miss AE classes with high frequencies, but low contribution such as Anal and Abdominal Pain in R04 trial. Therefore, clinical input such as relatedness to treatment and severity of AEs should also be considered and several configurations of CA biplots should be investigated such that conclusions need to be double-checked with frequencies tables, which are also provided in the Shiny app in the R-package \textit{visae}.


\begin{backmatter}


\section*{Funding}
This work was supported in part by the National Cancer Institute of the NIH (1U01CA232859-01) (MAD, GG, ML, SK, MT, GY, PAG, AR); and NIH National Center for Advancing Translational Science UCLA CTSI (UL1 TR001881-01) (MAD, MT, AR). Additional funding included support from the NIH for the original trials (U10-CA180868, U10-CA180822, UG1-CA189867, U10-CA180888, U10-CA180820, and U10-CA180821).

\section*{Abbreviations}
\begin{itemize}
    \item[] AE: Adverse Event;
    \item[] CA: Correspondence Analysis;
    \item[] CTCAE: Common Terminology Criteria for Adverse Events
    \item[] CRAN: Comprehensive R Archive Network
    \item[] RCT: Randomized Clinical Trial;
    \item[] Cape: Capecitabine;
    \item[] 5-FU:  5-fluorouracil;
    \item[] Oxa: oxaliplatin;
    \item[] RT: radiation  therapy;
    \item[] CTC: Common Toxicity Criteria.
\end{itemize}

\section*{Availability of data and materials}
The data that support the findings of this study are available from NRG Oncology but restrictions apply to the availability of these data, which were used under license for the current study, and so are not publicly available. Requests for the data, however, can be made to NRG Oncology at https://www.nrgoncology.org/Resources/Ancillary-Projects-Data-Sharing-Application. 

\section*{Ethics approval and consent to participate}
Not applicable.

\section*{Competing interests}
The authors declare that they have no competing interests.

\section*{Consent for publication}
Not applicable.

\section*{Authors' contributions}
MAD, GG and AR have design the study; GY and SK have acquired the data; MAD and SK have analyzed the data; MAD, AR, PG and NLH have interpreted the results; MAD and ML have developed the software; MAD and GG have drafted the manuscript; AR, MT, NLH, PG revised the manuscript. All authors read and approved the final manuscript.




\clearpage

\section*{Figures}

\begin{figure}[!ht]
	\centering
		\includegraphics[scale = 0.325]{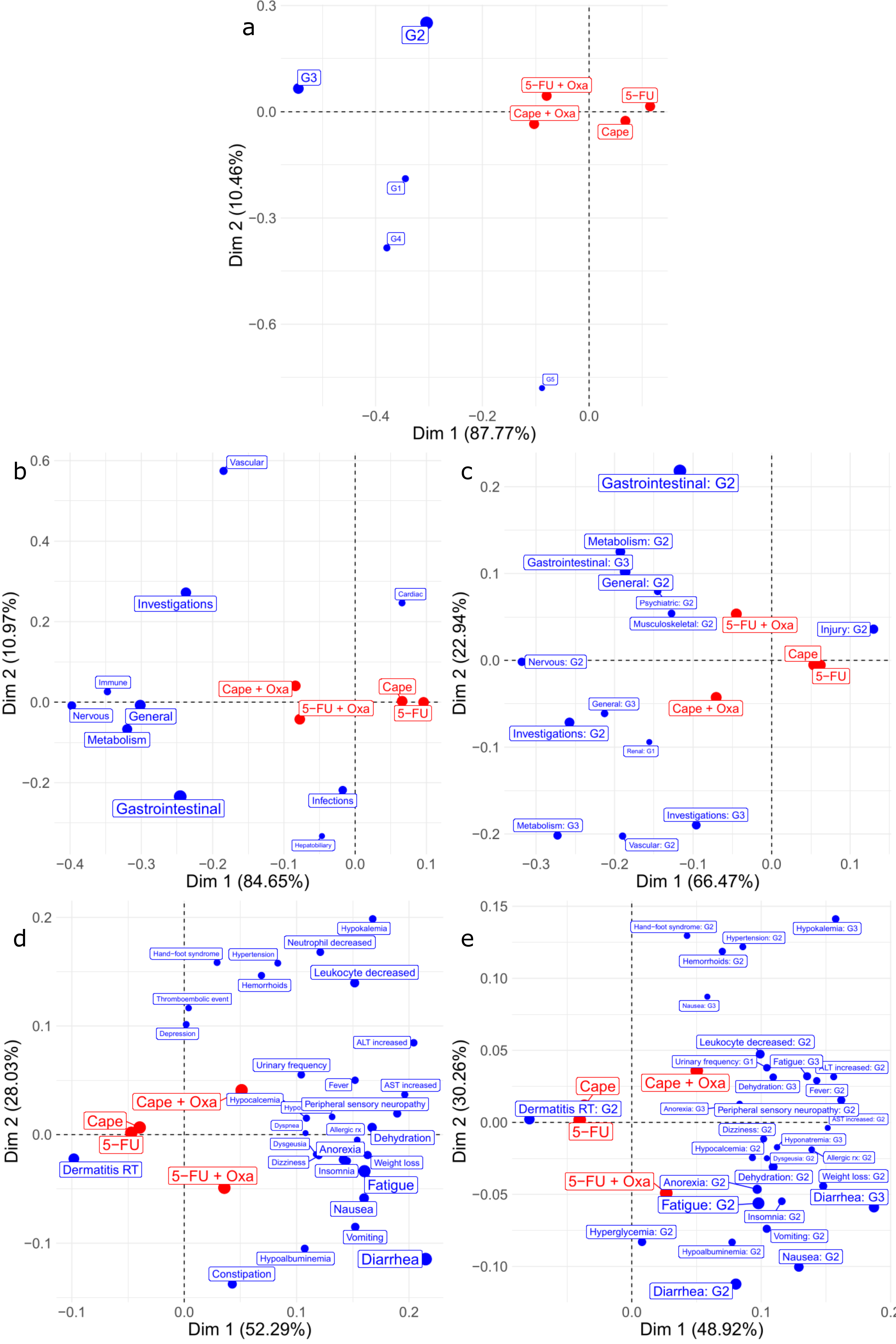}
    \caption{Asymmetric contribution biplots for AE data from R04 trial - a. AE class defined by AE grades; b. AE class defined by AE domains  with contribution at least 4.76\%; c. AE class defined by AE domains and grades with contribution and relative frequency at least 3.22\%; d. AE class defined by AE terms with contribution and relative frequency at least 0.96\%; e. AE class defined by AE terms and grades with contribution at least 0.64\% and  relative frequency at least 0.96\%.}
    \label{fig:r04_biplot}
\end{figure}

\newpage

\begin{figure}[!ht]
	\centering
		\includegraphics[scale = 0.325]{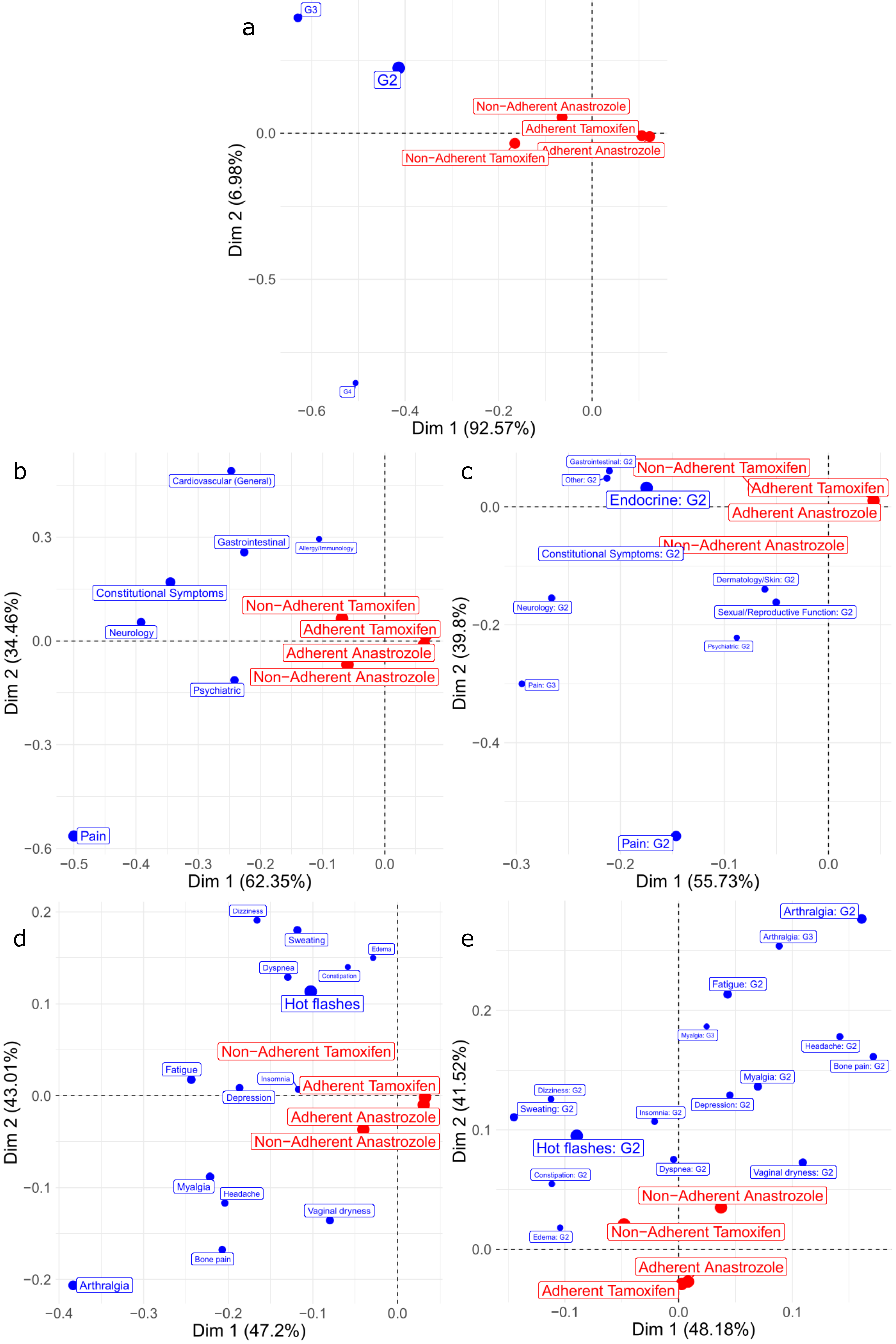}
    \caption{Asymmetric contribution biplots for AE at cycle 1 from B35 trial - a. AE class defined by AE grades; b. AE class defined by AE domains with contribution at least 5.0\%; c. AE class defined by AE domains and grades with contribution and relative frequency at least 1.64\%; d. AE class defined by AE terms with contribution at least 0.94\% and relative frequency at least 0.47\%; e. AE class defined by AE terms and grades with contribution and relative frequency at least 0.52\%.}
    \label{fig:b35_biplot}
\end{figure}



\clearpage


\bibliographystyle{bmc-mathphys} 
\bibliography{bmc_article}

\end{backmatter}
\end{document}


\begin{frontmatter}

\begin{fmbox}
\dochead{Research}


\title{Visualizing adverse events in clinical trials using correspondence analysis with R-package visae - Additional file}


\author[addressref={aff1}, 
  corref={aff1},            
  email={marcio.diniz@cshs.org}   
]{\inits{MAD}\fnm{M{\'a}rcio A.} \snm{Diniz}}
\author[
  addressref={aff1},
  email={gillian.gresham@cshs.org}
]{\inits{GG}\fnm{Gillian} \snm{Gresham}}
\author[
  addressref={aff1},
  email={sungjim.kim@cshs.org}
]{\inits{SK}\fnm{Sungjin} \snm{Kim}}
\author[
  addressref={aff1},
  email={michael.luu@cshs.org}
]{\inits{ML}\fnm{Michael} \snm{Luu}}
\author[
  addressref={aff2},
  email={norahh@med.umich.edu}
]{\inits{NLH}\fnm{N. Lynn} \snm{Henry}}
\author[
  addressref={aff1},
  email={mourad.tighiouart@cshs.org}
]{\inits{MT}\fnm{Mourad} \snm{Tighiouart}}
\author[
  addressref={aff3},
  email={yothersg@nrgoncology.org}
]{\inits{GY}\fnm{Greg} \snm{Yothers}}
\author[
  addressref={aff4},
  email={pganz@mednet.ucla.edu}
]{\inits{PG}\fnm{Patricia A.} \snm{Ganz}}
\author[
  addressref={aff1},
  email={andre.rogatko@cshs.org}
]{\inits{AR}\fnm{Andr{\'e}} \snm{Rogatko}}


\address[id=aff1]{
  \orgdiv{Samuel Oschin Compreheensive Cancer Center},             
  \orgname{Cedars-Sinai Medical Center},          
  \city{Los Angeles, CA},                              
  \cny{US}                                    
}

\address[id=aff2]{
  \orgdiv{Rogel Cancer Center},             
  \orgname{University of Michigan},          
  \city{Ann Arbor, MI},                              
  \cny{US}                                    
}

\address[id=aff3]{
  \orgdiv{Graduate School of Public Health},             
  \orgname{University of Pittsburgh and NRG Oncology},          
  \city{Pittsburgh, PA},                              
  \cny{US}                                    
}

\address[id=aff4]{
  \orgdiv{Fielding School of Public Health},             
  \orgname{UCLA Fielding School of Public Health},          
  \city{Los Angeles, CA},                              
  \cny{US}                                    
}



\end{fmbox}


\end{frontmatter}

\beginsupplement

\section*{Mathematical Description}
The goal of CA is to graphically represent contingency tables. In particular, for AE data, we consider a contingency table of $I$ AE classes and $J$ treatments as shown in Table \ref{tab:contingency_table01}. For now, AE class will be a generic label to indicate a variable that classifies AE into categories following a given criteria. 

\begin{table}[!ht]
\centering
\begin{tabular}{ccccc}
\toprule
\multirow{2}{*}{AE class} & \multicolumn{3}{c}{Treatment (T)} & 
\multirow{2}{*}{Total per AE}\\
                     & 1 & $\ldots$ &  J & \\
\midrule
1                    & $n_{11}$ & $\ldots$ & $n_{1J}$ & $n_{1.}$\\
\hline
$\vdots$             & $\vdots$ & $\vdots$ & $\vdots$ & $\vdots$ \\
\hline
I                    & $n_{I1}$ & $\ldots$ & $n_{IJ}$ & $n_{I.}$\\
\hline
Total per treatment      & $n_{.1}$ & $\ldots$ & $n_{.J}$ & $n_{..}$\\
\bottomrule
\end{tabular}
\caption{Contingency table $I \times J$ with row and column marginals}
\label{tab:contingency_table01}
\end{table}

The next step would be to the define treatment profiles as relative frequencies, $p_{ij} = n_{ij}/n_{.j}$ for $i = 1, \ldots, I$ and $j = 1,\ldots, J$. However, the column marginals in Table \ref{tab:contingency_table01} are the total number of AE instead of the total number of patients, and investigators are interested in contingency tables with the patient as the sample unit. Therefore, we need to redefine the contingency table before proceeding to apply correspondence analysis. First, we calculate a relative frequency table for each AE class $i$ given the total number of patients in each treatment as defined in Table \ref{tab:contingency_table02}.

\begin{table}[!ht]
\centering
\begin{tabular}{ccccc}
\toprule
\multirow{2}{*}{AE class} & \multicolumn{3}{c}{Treatment (T)} & 
\multirow{2}{*}{Total per AE}\\
                     & 1 & $\ldots$ &  J & \\
\midrule
i                     & $\pi_{i1}$ & $\ldots$ & $\pi_{ij}$ & $\sum_{j = 1}^J \pi_{ij}$\\
\hline
$i^C$                    & $1 - \pi_{i1}$ & $\ldots$ & $1 - \pi_{iJ}$ & $J - \sum_{j = 1}^J\pi_{ij}$\\
\hline
Total per treatment      & 1 & $\ldots$ & 1 & J\\
\bottomrule
\end{tabular}
\caption{Contingency table $2 \times J$ with row and column marginals}
\label{tab:contingency_table02}
\end{table}

\noindent where $\pi_{ij} = \sum_{l=1}^{N_{j}}I_{(n_{ijl} > 0)}/N_{j}$ is the relative frequency of $AE_i$ class for treatment $T_j$, $n_{ijl}$ is the frequency of AE class $i$  for patient $l$ receiving treatment $T_j$, and $N_{j}$ is the total number of patients receiving treatment $T_j$  for $i = 1, \ldots, I$ and $j = 1, \ldots, J$. Then, AE class specific contingency tables such as Table \ref{tab:contingency_table02} can be stacked generating an extended table for all AE classes. 

Following Greenacre \cite{greenacre2017}, we will apply CA on stacked tables such as Table \ref{tab:contingency_table03}. 

\begin{table}[!ht]
\centering
\begin{tabular}{ccccc}
\toprule
\multirow{2}{*}{AE class} & \multicolumn{3}{c}{Treatment (T)} & 
\multirow{2}{*}{Total per AE}\\
                     & 1 & $\ldots$ &  J & \\
\midrule
1                     & $\pi_{11}$ & $\ldots$ & $\pi_{1J}$ & $\sum_{j = 1}^J\pi_{1j}$\\
\hline
$1^c$                    & $1 - \pi_{11}$ & $\ldots$ & $1 - \pi_{1J}$ & $J - \sum_{j = 1}^J\pi_{1j}$\\
\hline
$\vdots$             & $\vdots$ & $\vdots$ & $\vdots$ & $\vdots$ \\
\hline
I                     & $\pi_{I1}$ & $\ldots$ & $\pi_{IJ}$ & $\sum_{j = 1}^J\pi_{Ij}$\\
\hline
$I^c$                    & $1 - \pi_{I1}$ & $\ldots$ & $1 - \pi_{IJ}$ & $J - \sum_{j = 1}^J\pi_{Ij}$\\
\hline
Total per treatment      & I & $\ldots$ & I & IJ\\
\bottomrule
\end{tabular}
\caption{Contingency table $2I \times J$ with row and column marginals}
\label{tab:contingency_table03}
\end{table}

\newpage 

We define column profiles as relative frequencies for each treatment,
\begin{eqnarray}
\mathbf{p_{.j}} &=& \begin{bmatrix} p_{ij} \\ p_{i^cj}\end{bmatrix}_{i=1}^{I} 
= \begin{bmatrix} \pi_{ij}/I \\ \pi_{i^cj}/I\end{bmatrix}_{i=1}^{I} 
= \begin{bmatrix} \pi_{ij}/I \\ (1 - \pi_{ij})/I\end{bmatrix}_{i=1}^{I} 
\label{eq:column_profiles}
\end{eqnarray}
resulting in a $2I \times J$ correspondence matrix $\mathbf{P}= [p_{.j}]_{j = 1}^J$ associated with Table \ref{tab:contingency_table03}. Although CA can be based either on column or row profiles, we will focus only on the analysis for column profiles. Both analyses are mathematically equivalent, but they lead us to different interpretations. The interpretation of Table \ref{tab:contingency_table03} is asymmetric: we are interested in studying the differences in treatment profiles that lie in the space generated by the column profiles, which will be denoted as toxicity space.

The toxicity space has dimension $K = min\{I, J\}$, such that the vertices of such space are given by extreme treatment profiles, i.e., treatment profiles that have one as a relative frequency for AE class $i$ and zero for all other AE classes for $i = 1, \ldots, I$. The toxicity space can be understood using summary measures of location and dispersion as follows. First, we calculate column marginal relative frequencies, denoted as masses in CA,
\begin{eqnarray}
\mathbf{c}_{J \times 1} &=& [c_j]_{j = 1}^J = \left[\frac{I}{IJ}\right]_{j = 1}^J = \left[\frac{1}{J}\right]_{j = 1}^J
\label{eq:treatment_masses}
\end{eqnarray}

The masses associated with each treatment are the same, independent of the number of patients or number of AEs, which is desirable because there is no particular reason to give a higher weight for a specific group in our analysis. Although treatments are often equally randomized in clinical trials, we also can define groups as treatment adherent or non-adherent which will not have the same number of patients as will be discussed in the next section.

Then, the average toxicity profile is defined as the row marginal relative frequencies. It also can be calculated as a weighted average of treatment profiles (\ref{eq:column_profiles}) with weights given by the treatment masses (\ref{eq:treatment_masses}),
\begin{eqnarray}
\mathbf{r}_{2I \times 1} &=& \begin{bmatrix} r_i \\ r_{i^c}\end{bmatrix}_{i=1}^{I}  = \sum_{j = 1}^J c_j p_{.j} \nonumber \\ 
&=& \begin{bmatrix} \bar{\pi}_{i.}/I \\ (1 - \bar{\pi}_{i.})/I\end{bmatrix}_{i=1}^{I},
\label{eq:average}
\end{eqnarray}
where $\bar{\pi}_{i.} = \frac{1}{J}\sum_{j = 1}^J \pi_{ij}$.

In CA, the distance of treatment profiles from the average profile is based on a weighted Euclidean distance known as the $\chi^2$-distance:
\begin{eqnarray}
\mbox{$\chi^2$-distance}(\mathbf{p_{.j}}, \mathbf{r}) = \sqrt{\frac{(p_{1j} - r_1)^2}{r_1} + \ldots + \frac{(p_{I^cj} - r_{I^c})^2}{r_{I^c}}}.
\label{eq:chidistance}
\end{eqnarray}

Then, we can calculate the variability of treatment profiles as a weighted average $\chi^2$ distance \eqref{eq:chidistance} of treatment profiles from the average, with weights given by their treatment masses (\ref{eq:treatment_masses}). This is known as total inertia in CA and is given by
\begin{eqnarray}
\mbox{total inertia} &=& \sum_{j = 1}^J c_j \mbox{$\chi^2$-distance}(p_{.j}, \mathbf{r})^2 \nonumber\\
&=& \frac{1}{IJ} \sum_{j = 1}^J \sum_{i = 1}^I \frac{(\pi_{ij} - \bar{\pi}_{i.})^2}{ \bar{\pi}_{i.}(1 -  \bar{\pi}_{i.})}
\label{eq:total_inertia}
\end{eqnarray}
which can be interpreted as the average total inertia for all AE classes.
Therefore, any toxicity space can be summarized by the average \eqref{eq:average} and variance \eqref{eq:total_inertia} as in any statistical problem. As a next step, we can standardize the correspondence matrix $\mathbf{P}$, 
\begin{eqnarray}
\mathbf{R} &=& \mathbf{D}_c^{-1/2}(\mathbf{P} - \mathbf{c}\mathbf{r}^T)\mathbf{D}_r^{-1/2}\nonumber\\
&=& \mathbf{D}_c^{1/2}(\mathbf{D}_c^{-1}\mathbf{P} - \mathbf{1}\mathbf{r}^T)\mathbf{D}_r^{-1/2}
\label{eq:standardized}
\end{eqnarray}
where $\mathbf{D}_c$ and $\mathbf{D}_r$ are $J \times J$ and $2I \times 2I $ diagonal matrices defined based on \eqref{eq:treatment_masses} and \eqref{eq:average}, respectively. The standardized residual treatment profiles \eqref{eq:standardized} represent the differences in comparison to the average treatment profile assuming that groups were identical. 

Visualizing the standardized residuals treatment profiles in the toxicity space can give us insight regarding the association between treatments and AE classes. Nonetheless, it is not always feasible to display residual treatment profiles when the number of dimensions is greater than three, i.e., four treatment arms $(J \geq 4)$ or four AE classes ($I \geq 4$). Moreover, distances between standardized residual treatment profiles are not simple to be evaluated even in a three dimensional space. 

In this context, CA seeks the two-dimensional display that minimizes the weighted sum of $\chi^2$-distance (\ref{eq:chidistance}) between the residual treatment profiles that lie in the $K$-dimensional toxicity space and their projection on a two-dimensional display candidate, where the weights are given by the treatment masses \eqref{eq:treatment_masses}. The solution for this minimization problem is given by biplots \cite{greenacre1993}. The biplot for CA is defined based on the singular value decomposition of \eqref{eq:standardized}:
\begin{eqnarray}
\mathbf{R} &=& \mathbf{U}\mathbf{D}_\alpha\mathbf{V}^T
\end{eqnarray}
where $\mathbf{U}\mathbf{U}^T = \mathbf{V}^T\mathbf{V} = \mathbf{I}_K$, such that $\mathbf{U}$ and $\mathbf{V}$ are  $J \times K$ and $2I \times K$ matrices, respectively, with each column in both matrices representing the dimension $k$ of the toxicity space for $k = 1, \ldots, K$. Furthermore, $\mathbf{D}_\alpha$ is the diagonal matrix of single values,
\begin{eqnarray}
\mathbf{D}_\alpha = diag(\alpha_1 \geq \ldots \geq \alpha_K)
\end{eqnarray}
such that $\sum_{k = 1}^K \alpha_k^2$ is the total inertia given in (\ref{eq:total_inertia}); $\mathbf{D}_c, \mathbf{D}_r$ are diagonal matrices with diagonals given by (\ref{eq:treatment_masses}) and (\ref{eq:average}), respectively. Then, the asymmetric contribution biplot proposed by \cite{greenacre2013} will display two sets of dots: (a) dots representing treatment profiles with coordinates given by $\mathbf{F} = \mathbf{D}_c^{-1/2}\mathbf{U}\mathbf{D}_\alpha$, which are denoted as principal coordinates; and dots representing AE classes with coordinates given by $\mathbf{V}$, such that $v_{ik}^2$ is the contribution of AE class $i$ in dimension $k$ for $k = 1, \ldots, K$ with $\sum_{i = 1}^I v_{ik}^2 = \alpha_k^2$. The expected contribution of an AE class is the average contribution assuming that all AE classes have the same contributions, i.e., $1/I$; the expected mass of an class is the average frequency assuming that all AE classes have the same frequency, i.e., $1/J$.

Biplots show the main features of high dimensional data using only two dimensions $(K = 2)$ while minimizing the loss of information. The first dimension of the biplot represents the direction with highest inertia ($\alpha_1^2$) in the toxicity space and the second dimension corresponds to the direction with the second highest inertia $(\alpha_2^2)$. Adding up the inertia of the remaining dimensions $(\sum_{k = 3}^K \alpha_k^2)$ allows us to quantify the loss of information of projecting the toxicity space into a two-dimensional plane. The inertia associated with each dimension $(\alpha_k^2)$ can also be broken down in contributions $(v_{ik}^2)$ of each AE class for $i = 1, \ldots, I$ and $k = 1, \ldots, K$. Therefore, we are able to compare treatment profiles based on the position of their dots relative to the origin, which represents the average treatment (\ref{eq:average}), and interpret each axis based on the position of dots associated with each AE class. 

Finally, AE classes can be defined in Table \ref{tab:contingency_table03} based on three levels of data aggregation: (a) AE grades, (b) AE domains, (c) AE terms and their combinations. While toxicity profiles can be presented and compared based on tables when AE classes are only defined by AE grades, it is a much more complex task when AE classes are defined by AE domains or AE terms, and their combinations with AE grades. There are 26 domains and 790 AE terms in CTCAE v4, which can generate 130 AE classes when AE domains are combined with AE grades and 3950 AE classes when AE terms are combined with AE grades.


\newpage

\section*{Figures}

\begin{figure}[!h]
	\centering
		\includegraphics[scale = 0.325]{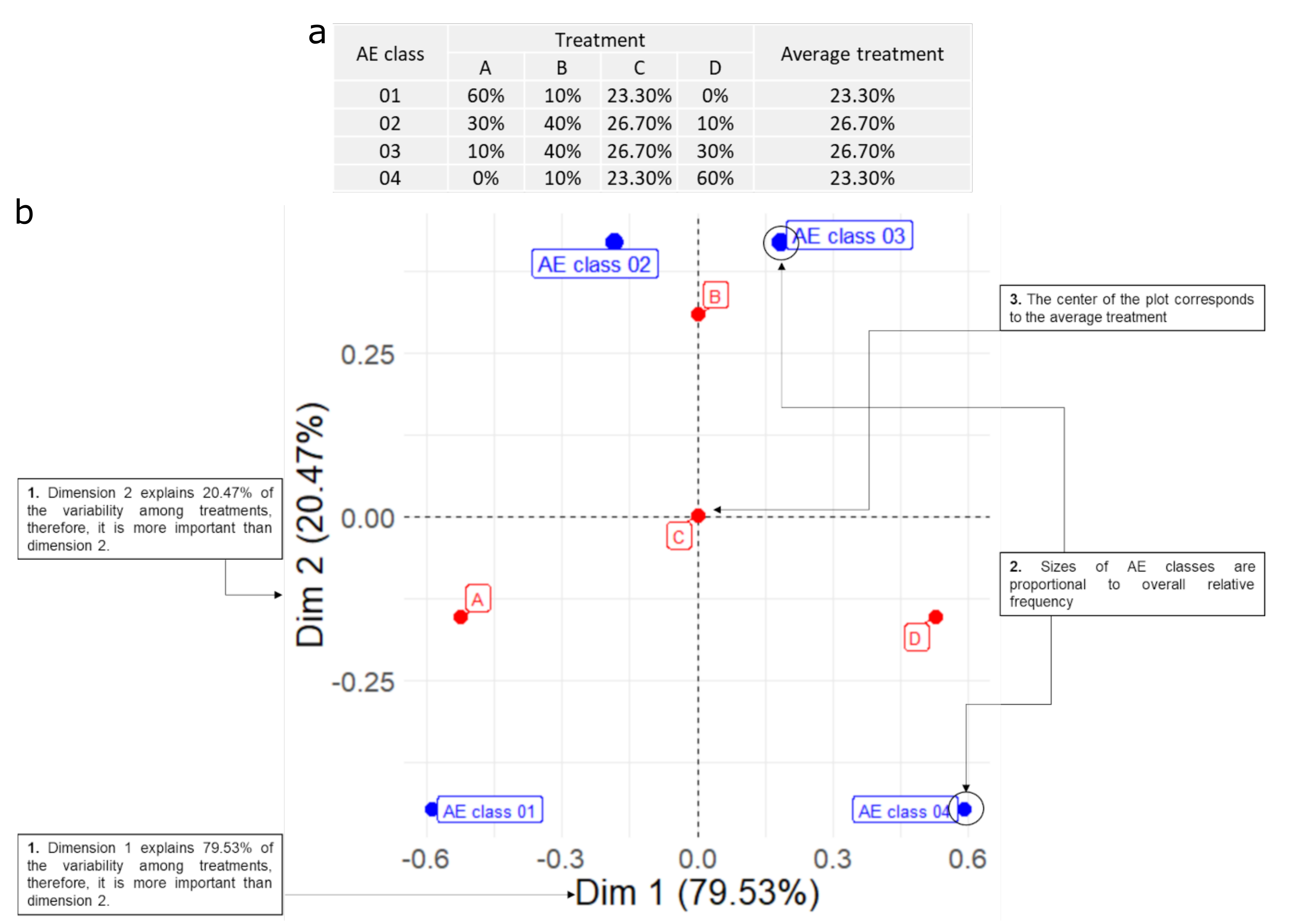}
    \caption{Asymmetric contribution biplot for toy data - a. Contingency table of AE classes by treatment; b. Interpreting variance explained by dimensions, the size of AE classes (blue dots) and the center of CA biplot. }
    \label{fig:r04_biplot}
\end{figure}

\begin{figure}[!h]
	\centering
		\includegraphics[scale = 0.325]{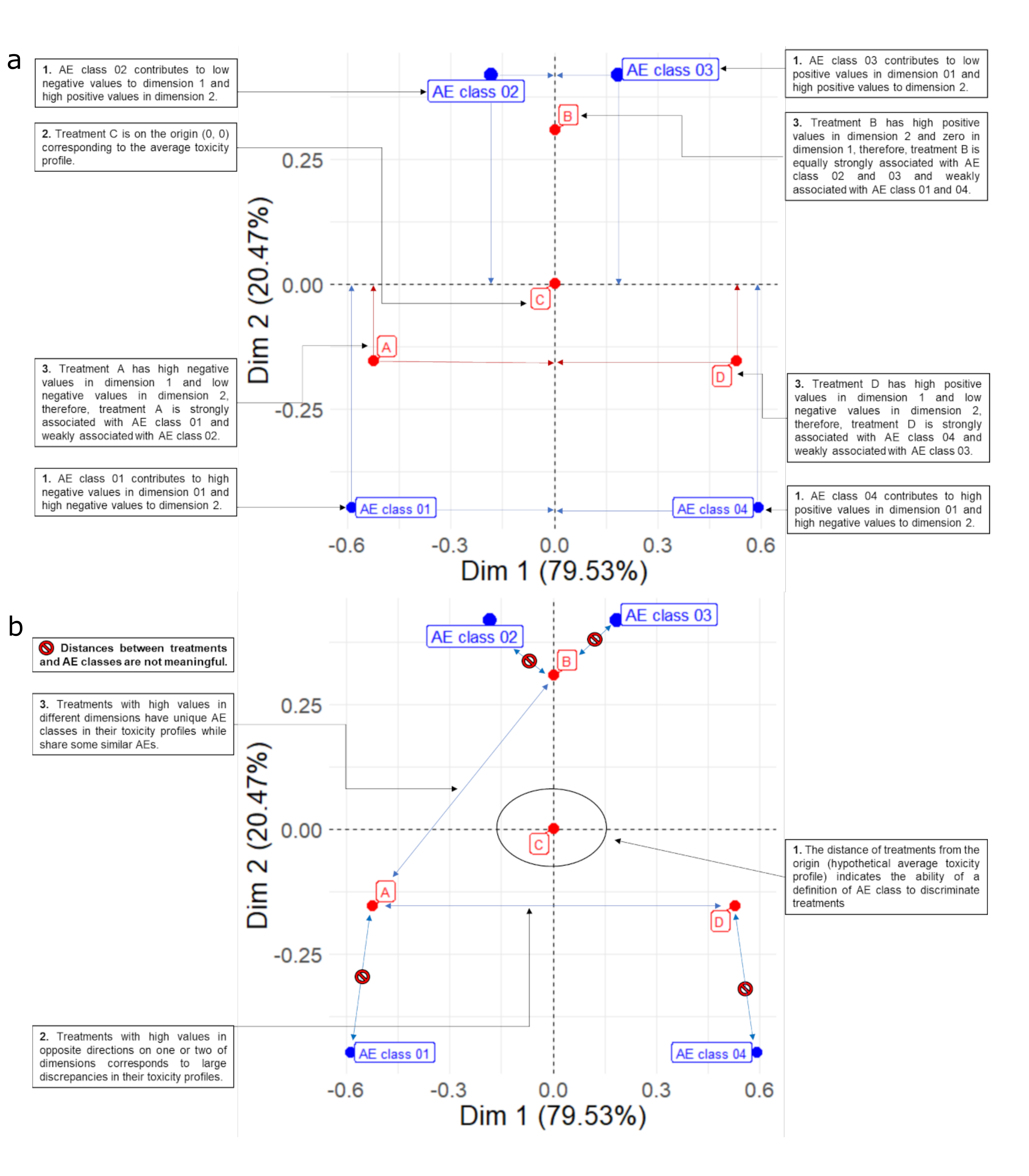}
    \caption{Asymmetric contribution biplot for toy data - a. Interpreting AE classes (blue dots) and groups (red dots) relative to dimension 1 and dimension 2; b. Interpreting the distances among groups (red dots).}
    \label{fig:r04_biplot}
\end{figure}

\clearpage

\section*{Tables}


\subsection*{R04}

\begin{table}[!ht]
    \centering
    \begin{tabular}{c|l|l|l|l|l}
    \toprule
        \multirow{2}{*}{Domain} & 5-FU  & 5-FU + Oxa & Cape & Cape + Oxa  & \multirow{2}{*}{Average} \\ 
            & (n = 328) & (n = 327) & (n = 325) &  (n = 328) &  \\ \hline
        Cardiac & 0.92 & 0.31 & 1.85 & 1.22 & 1.07 \\ \hline
        Gastrointestinal & 42.68 & 60.86 & 47.08 & 54.88 & 51.37 \\ \hline
        General & 19.51 & 33.33 & 24.31 & 33.54 & 27.67 \\ \hline
        Hepatobiliary & 0.00 & 0.61 & 0.31 & 0.00 & 0.23 \\ \hline
        Immune & 0.00 & 3.67 & 0.31 & 3.96 & 1.99 \\ \hline
        Infections & 7.01 & 8.26 & 6.46 & 6.10 & 6.96 \\ \hline
        Investigations & 19.51 & 24.46 & 17.54 & 29.88 & 22.85 \\ \hline
        Metabolism & 13.41 & 24.77 & 14.46 & 24.09 & 19.18 \\ \hline
        Nervous & 3.96 & 13.15 & 4.92 & 13.41 & 8.86 \\ \hline
        Vascular & 3.35 & 4.28 & 4.31 & 9.45 & 5.35 \\  
        \bottomrule
    \end{tabular}
\caption{Percentage (\%) of selected AE domains  with contribution at least 4.76\% by treatment in R04 trial}
\label{tab:r04_ae_domain_percentage}
\end{table}

\begin{table}[!ht]
    \centering
    \begin{tabular}{c|l|l|l|l|l}
    \toprule
        \multirow{2}{*}{Domain:Grade} & 5-FU  & 5-FU + Oxa & Cape & Cape + Oxa  & \multirow{2}{*}{Average} \\ 
            & (n = 328) & (n = 327) & (n = 325) &  (n = 328) &  \\ \hline
        Gastrointestinal: G2 & 38.72 & 52.60 & 40.92 & 43.29 & 43.88 \\ \hline
        Gastrointestinal: G3 & 10.37 & 23.55 & 14.15 & 22.26 & 17.58 \\ \hline
        General: G2 & 17.68 & 28.75 & 20.92 & 26.83 & 23.55 \\ \hline
        General: G3 & 1.83 & 5.20 & 2.77 & 7.01 & 4.20 \\ \hline
        Injury: G2 & 14.63 & 11.31 & 14.15 & 9.45 & 12.39 \\ \hline
        Investigations: G2 & 12.50 & 18.35 & 10.15 & 22.56 & 15.89 \\ \hline
        Investigations: G3 & 7.62 & 7.64 & 8.31 & 12.50 & 9.02 \\ \hline
        Metabolism: G2 & 11.59 & 20.49 & 12.62 & 18.29 & 15.75 \\ \hline
        Metabolism: G3 & 3.96 & 7.34 & 4.00 & 12.80 & 7.03 \\ \hline
        Musculoskeletal: G2 & 3.35 & 6.12 & 3.38 & 5.79 & 4.66 \\ \hline
        Nervous: G2 & 3.35 & 10.70 & 4.00 & 12.50 & 7.64 \\ \hline
        Psychiatric: G2 & 1.83 & 5.81 & 3.69 & 5.18 & 4.13 \\ \hline
        Renal: G1 & 1.22 & 2.45 & 1.23 & 3.96 & 2.21 \\ \hline
        Vascular: G2 & 3.35 & 3.36 & 0.92 & 7.01 & 3.66 \\ 
    \bottomrule
    \end{tabular}
\caption{Percentage (\%) of selected AE domain:grade classes  with contribution and mass at least 3.22\% by treatment in R04 trial}
\label{tab:r04_ae_domain_grade_percentage}
\end{table}

\begin{table}[!ht]
    \centering
    \begin{tabular}{c|l|l|l|l|l}
    \toprule
        \multirow{2}{*}{Term} & 5-FU  & 5-FU+Oxa & Cape & Cape+Oxa  & \multirow{2}{*}{Average} \\ 
            & (n = 328) & (n = 327) & (n = 325) &  (n = 328) &  \\ \hline
          \makecell{Alanine \\aminotransferase \\increased} & 0.61 &   3.06 & 0.31 &   5.49 & 2.37 \\ \hline
        Allergic reaction & 0.00 & 2.14 & 0.31 & 2.44 & 1.22 \\ \hline
        Anorexia & 4.57 & 10.40 & 6.15 & 10.37 & 7.87 \\ \hline
        Aspartate \\aminotransferase \\increased & 0.00 & 2.75 & 0.31 & 3.96 & 1.76 \\ \hline
        Constipation & 3.96 & 9.79 & 7.69 & 5.18 & 6.66 \\ \hline
        Dehydration & 3.96 & 10.09 & 5.85 & 11.28 & 7.79 \\ \hline
        Depression & 0.92 & 0.61 & 2.15 & 2.13 & 1.45 \\ \hline
        Radiation dermatitis & 16.46 & 12.54 & 15.69 & 10.67 & 13.84 \\ \hline
        Diarrhea & 18.60 & 36.70 & 20.31 & 31.40 & 26.75 \\ \hline
        Dizziness & 0.61 & 3.06 & 1.23 & 3.05 & 1.99 \\ \hline
        Dysgeusia & 0.00 & 1.83 & 0.61 & 1.83 & 1.07 \\ \hline
        Dyspnea & 0.61 & 1.53 & 0.00 & 1.83 & 0.99 \\ \hline
        Fatigue & 15.55 & 28.44 & 21.23 & 27.74 & 23.24 \\ \hline
        Fever & 1.22 & 2.75 & 0.31 & 4.27 & 2.14 \\ \hline
        Hemorrhoids & 0.92 & 0.92 & 1.85 & 3.66 & 1.83 \\ \hline
        Hypertension & 1.22 & 0.61 & 0.61 & 3.35 & 1.45 \\ \hline
        Hypoalbuminemia & 2.44 & 5.20 & 0.92 & 3.35 & 2.98 \\ \hline
        Hypocalcemia & 0.92 & 2.45 & 0.92 & 3.05 & 1.83 \\ \hline
        Hypokalemia & 0.92 & 1.53 & 0.92 & 6.10 & 2.37 \\ \hline
        Hyponatremia & 0.61 & 1.83 & 0.00 & 2.44 & 1.22 \\ \hline
        Insomnia & 0.92 & 4.28 & 1.54 & 4.27 & 2.75 \\ \hline
        Nausea & 5.49 & 12.84 & 6.46 & 11.59 & 9.10 \\ \hline
        \makecell{Neutrophil count\\ decreased} & 0.61 & 1.83 & 2.77 & 5.79 & 2.75 \\ \hline
       \makecell{ Palmar-plantar \\erythrodysesthesia \\syndrome} & 0.30 & 0.31 & 2.46 & 2.74 & 1.45 \\ \hline
        \makecell{Peripheral sensory \\neuropathy} & 0.61 & 5.20 & 2.15 & 6.40 & 3.59 \\ \hline
        Thromboembolic event & 0.30 & 0.31 & 2.15 & 1.83 & 1.15 \\ \hline
        Urinary frequency & 1.52 & 2.75 & 1.54 & 4.27 & 2.52 \\ \hline
        Vomiting & 2.74 & 7.34 & 1.85 & 5.79 & 4.43 \\ \hline
        Weight loss & 2.13 & 5.81 & 1.54 & 6.10 & 3.90 \\ \hline
        \makecell{White blood cell\\ decreased} & 4.27 & 5.81 & 4.00 & 11.28 & 6.34 \\ 
    \bottomrule
    \end{tabular}
    \caption{Percentage (\%) of selected AE terms  with contribution and mass at least 0.96\% by treatment in R04 trial}
\label{tab:r04_ae_term_percentage}
\end{table}

\begin{table}[!ht]
    \centering
    \begin{tabular}{c|l|l|l|l|l}
    \toprule
        \multirow{2}{*}{Term:Grade} & 5-FU  & 5-FU+Oxa & Cape & Cape+Oxa  & \multirow{2}{*}{Average} \\ 
            & (n = 328) & (n = 327) & (n = 325) &  (n = 328) &  \\ \hline
        \makecell{Alanine \\aminotransferase \\increased: G2} & 0.61 & 2.14 & 0.00 & 3.66 & 1.60 \\ \hline
        Allergic reaction: G2 & 0.00 & 2.14 & 0.31 & 2.44 & 1.22 \\ \hline
        Anorexia: G2 & 3.96 & 8.87 & 5.54 & 7.93 & 6.57 \\ \hline
        Anorexia: G3 & 0.61 & 1.53 & 0.61 & 2.13 & 1.22 \\ \hline
        \makecell{Aspartate \\aminotransferase \\increased: G2} & 0.00 & 1.83 & 0.00 & 2.44 & 1.07 \\ \hline
        Dehydration: G2 & 3.66 & 7.34 & 3.69 & 7.32 & 5.50 \\ \hline
        Dehydration: G3 & 0.30 & 2.75 & 2.15 & 3.96 & 2.29 \\ \hline
        \makecell{Radiation \\Dermatitis: G2} & 14.02 & 10.40 & 13.23 & 9.45 & 11.78 \\ \hline
        Diarrhea: G2 & 11.89 & 20.80 & 13.54 & 15.24 & 15.37 \\ \hline
        Diarrhea: G3 & 6.71 & 15.90 & 6.77 & 15.55 & 11.23 \\ \hline
        Dizziness: G2 & 0.61 & 2.45 & 0.92 & 2.74 & 1.68 \\ \hline
        Dysgeusia: G2 & 0.00 & 1.83 & 0.61 & 1.83 & 1.07 \\ \hline
        Fatigue: G2 & 14.33 & 24.46 & 19.08 & 21.95 & 19.96 \\ \hline
        Fatigue: G3 & 1.22 & 3.98 & 2.15 & 5.79 & 3.29 \\ \hline
        Fever: G2 & 0.92 & 2.45 & 0.31 & 3.96 & 1.91 \\ \hline
        Hemorrhoids: G2 & 0.92 & 0.92 & 1.85 & 3.66 & 1.83 \\ \hline
        Hyperglycemia: G2 & 3.66 & 5.50 & 3.69 & 3.05 & 3.98 \\ \hline
        Hypertension: G2 & 1.22 & 0.61 & 0.61 & 3.35 & 1.45 \\ \hline
        Hypoalbuminemia: G2 & 1.83 & 3.67 & 0.61 & 2.44 & 2.14 \\ \hline
        Hypocalcemia: G2 & 0.61 & 2.45 & 0.92 & 2.44 & 1.60 \\ \hline
        Hypokalemia: G3 & 0.92 & 1.53 & 0.92 & 5.79 & 2.29 \\ \hline
        Hyponatremia: G3 & 0.61 & 1.83 & 0.00 & 2.13 & 1.14 \\ \hline
        Insomnia: G2 & 0.92 & 4.28 & 1.54 & 3.66 & 2.60 \\ \hline
        Nausea: G2 & 5.18 & 12.23 & 5.23 & 9.45 & 8.02 \\ \hline
        Nausea: G3 & 0.30 & 0.61 & 1.23 & 2.13 & 1.07 \\ \hline
        \makecell{Palmar-plantar\\ erythrodysesthesia \\syndrome: G2} & 0.00 & 0.31 & 2.15 & 2.44 & 1.23 \\ \hline
        \makecell{Peripheral \\sensory \\neuropathy: G2} & 0.61 & 4.59 & 2.15 & 6.10 & 3.36 \\ \hline
        Urinary frequency: G1 & 1.22 & 2.45 & 1.23 & 3.96 & 2.21 \\ \hline
        Vomiting: G2 & 2.44 & 5.81 & 1.85 & 4.57 & 3.67 \\ \hline
        Weight loss: G2 & 1.83 & 5.50 & 1.23 & 5.49 & 3.51 \\ \hline
        \makecell{White blood cell \\decreased: G2} & 4.27 & 5.50 & 3.69 & 8.23 & 5.42 \\
    \bottomrule
    \end{tabular}
        \caption{Percentage (\%) of selected AE term:grade classes with contribution at least 0.64\% and  mass at least 0.96\% by treatment in R04 trial}
\label{tab:r04_ae_term_grade_percentage}
\end{table}

\clearpage
\subsection*{B35}

\begin{table}[!ht]
    \centering
    \begin{tabular}{c|l|l|l|l|l}
    \toprule
        \multirow{3}{*}{Domain} & \multicolumn{2}{|c|}{Adherent} & \multicolumn{2}{|c|}{Non-adherent} & \multirow{3}{*}{Average} \\

                                & Anastrozole & Tamoxifen &  Anastrozole & Tamoxifen &  \\
                                &  (n = 1065) & (n = 1074) & (n = 443) & (n = 427) &  \\
        \hline
        Allergy/Immunology & 0.19 & 0.28 & 0.00 & 1.17 & 0.41 \\ \hline
        Cardiovascular & 1.41 & 1.58 & 1.13 & 6.09 & 2.55 \\ \hline
        \makecell{Constitutional\\ Symptoms} & 5.82 & 6.24 & 10.16 & 13.58 & 8.95 \\ \hline
        Gastrointestinal & 1.88 & 2.70 & 3.16 & 6.09 & 3.46 \\ \hline
        Neurology & 2.44 & 1.58 & 6.09 & 7.26 & 4.34 \\ \hline
        Pain & 11.36 & 8.38 & 27.99 & 15.46 & 15.80 \\ \hline
        Psychiatric & 1.78 & 2.23 & 5.19 & 3.98 & 3.30 \\ 
    \bottomrule
    \end{tabular}
    \caption{Percentage (\%) of selected AE domains at cycle 1 with contribution at least 5.0\% by treatment in B35 trial}
    \label{tab:b35_ae_domain_percentage}
\end{table}

\begin{table}[!ht]
    \centering
    \begin{tabular}{c|l|l|l|l|l}
    \toprule
        \multirow{3}{*}{Domain} & \multicolumn{2}{|c|}{Adherent} & \multicolumn{2}{|c|}{Non-adherent} & \multirow{3}{*}{Average} \\
 
                                & Anastrozole & Tamoxifen &  Anastrozole & Tamoxifen &  \\
                                &  (n = 1065) & (n = 1074) & (n = 443) & (n = 427) &  \\
        \hline
        \makecell{Constitutional\\ Symptoms: G2} & 5.73 & 6.05 & 10.16 & 11.94 & 8.47 \\ \hline
        \makecell{Dermatology/\\Skin: G2} & 2.82 & 2.42 & 4.51 & 3.28 & 3.26 \\ \hline
        Endocrine: G2 & 22.63 & 26.63 & 27.31 & 31.85 & 27.11 \\ \hline
        Gastrointestinal: G2 & 1.69 & 2.61 & 2.94 & 5.15 & 3.1 \\ \hline
        Neurology: G2 & 1.97 & 1.4 & 5.19 & 5.39 & 3.49 \\ \hline
        Other: G2 & 2.16 & 2.05 & 2.94 & 5.15 & 3.07 \\ \hline
        Pain: G2 & 10.8 & 7.63 & 23.25 & 11.71 & 13.35 \\ \hline
        Pain: G3 & 0.66 & 0.84 & 5.64 & 4.21 & 2.84 \\ \hline
        Psychiatric: G2 & 1.69 & 1.86 & 4.51 & 2.58 & 2.66 \\ \hline
        \makecell{Sexual/Reproductive\\ Function: G2} & 4.13 & 2.7 & 5.64 & 3.9 & 4.11 \\
    \bottomrule
    \end{tabular}
    \caption{Percentage (\%) of selected AE domain:grade classes at cycle 1 with contribution at least 1.64\% by treatment in B35 trial}
    \label{tab:b35_ae_domain_grade_percentage}
\end{table}

\begin{table}[!ht]
    \centering
    \begin{tabular}{c|l|l|l|l|l}
    \toprule
        \multirow{3}{*}{Term} & \multicolumn{2}{|c|}{Adherent} & \multicolumn{2}{|c|}{Non-adherent} & \multirow{3}{*}{Average} \\
 
                                & Anastrozole & Tamoxifen &  Anastrozole & Tamoxifen &  \\
                                &  (n = 1065) & (n = 1074) & (n = 443) & (n = 427) &  \\
        \hline
        Arthralgia& 7.04 & 4.66 & 18.51 & 9.6 & 9.95 \\ \hline
        Bone pain & 1.6 & 0.84 & 4.74 & 1.64 & 2.2 \\ \hline
        Constipation & 0.19 & 1.12 & 0.68 & 1.87 & 0.96 \\ \hline
        \makecell{Dizziness/\\lightheadedness} & 0.56 & 0.47 & 1.35 & 3.28 & 1.42 \\ \hline
        Dyspnea & 1.03 & 1.58 & 2.26 & 3.75 & 2.15 \\ \hline
        Edema & 0.66 & 0.84 & 0.45 & 1.87 & 0.95 \\ \hline
        Fatigue & 3.29 & 2.61 & 7.45 & 6.79 & 5.03 \\ \hline
        Headache & 0.75 & 1.02 & 3.84 & 1.64 & 1.81 \\ \hline
        Hot flashes/flushes & 22.54 & 26.54 & 27.09 & 31.62 & 26.94 \\ \hline
        Insomnia & 1.13 & 0.65 & 2.03 & 1.87 & 1.42 \\ \hline
        \makecell{Mood alteration/\\
        depression} & 1.41 & 1.86 & 4.29 & 3.75 & 2.83 \\ \hline
        Myalgia & 3.29 & 2.33 & 7.45 & 4.68 & 4.44 \\ \hline
        Radiation dermatitis & 1.22 & 0.56 & 0.9 & 1.64 & 1.08 \\ \hline
        Rash/desquamation & 2.16 & 2.33 & 3.84 & 2.34 & 2.67 \\ \hline
        Sweating & 2.07 & 2.89 & 3.16 & 6.09 & 3.55 \\ \hline
        Vaginal dryness & 3.85 & 1.96 & 4.97 & 2.34 & 3.28 \\ \bottomrule
    \end{tabular}
    \caption{Percentage (\%) of selected AE terms at cycle 1 with contribution at least 0.94\% and mass at least 0.47\%  by treatment in B35 trial}
     \label{tab:b35_ae_term_percentage}
\end{table}

\begin{table}[!ht]
    \centering
    \begin{tabular}{c|l|l|l|l|l}
    \toprule
        \multirow{3}{*}{Term:Grade} & \multicolumn{2}{|c|}{Adherent} & \multicolumn{2}{|c|}{Non-adherent} & \multirow{3}{*}{Average} \\
 
                                & Anastrozole & Tamoxifen &  Anastrozole & Tamoxifen &  \\
                                &  (n = 1065) & (n = 1074) & (n = 443) & (n = 427) &  \\
        \hline
        Arthralgia: G2 & 6.761 & 4.38 & 15.12 & 7.73 & 8.5 \\ \hline
        Arthralgia: G3 & 0.282 & 0.28 & 3.61 & 1.64 & 1.45 \\ \hline
        Bone pain: G2 & 1.596 & 0.74 & 4.29 & 0.94 & 1.89 \\ \hline
        Chest pain: G2 & 0.282 & 0.47 & 1.35 & 0.94 & 0.76 \\ \hline
        Constipation: G2 & 0.188 & 1.12 & 0.68 & 1.87 & 0.96 \\ \hline
        \makecell{Dizziness/\\lightheadedness: G2} & 0.563 & 0.37 & 1.13 & 2.34 & 1.1 \\ \hline
        Dyspareunia: G2 & 0.751 & 0.19 & 1.13 & 0.7 & 0.69 \\ \hline
        Dyspnea: G2 & 0.939 & 1.4 & 2.03 & 1.87 & 1.56 \\ \hline
        Dyspnea: G3 & 0.188 & 0.19 & 0.23 & 1.64 & 0.56 \\ \hline
        Edema: G2 & 0.657 & 0.84 & 0.45 & 1.64 & 0.9 \\ \hline
        Fatigue: G2 & 3.192 & 2.42 & 7.45 & 5.39 & 4.61 \\ \hline
        Headache: G2 & 0.469 & 1.02 & 3.61 & 0.94 & 1.51 \\ \hline
        \makecell{Hot flashes/\\flushes: G2} & 22.535 & 26.54 & 27.09 & 31.62 & 26.94 \\ \hline
        \makecell{Infection without \\neutropenia: G3} & 0.469 & 0.84 & 0 & 0.94 & 0.56 \\ \hline
        Insomnia: G2 & 0.939 & 0.65 & 1.81 & 1.87 & 1.32 \\ \hline
        \makecell{Mood alteration-\\depression: G2} & 1.315 & 1.68 & 3.61 & 2.34 & 2.24 \\ \hline
        \makecell{Mood alteration-\\depression: G3} & 0.094 & 0.19 & 0.45 & 1.4 & 0.53 \\ \hline
        Myalgia: G2 & 3.192 & 2.14 & 5.64 & 3.51 & 3.62 \\ \hline
        Myalgia : G3 & 0.094 & 0.19 & 1.81 & 1.17 & 0.81 \\ \hline
        Nausea: G2 & 0.376 & 0.19 & 1.13 & 1.17 & 0.71 \\ \hline
        \makecell{Neuropathy-\\sensory: G2} & 0.376 & 0.19 & 2.03 & 0.23 & 0.71 \\ \hline
        Sweating: G2 & 2.066 & 2.89 & 3.16 & 6.09 & 3.55 \\ \hline
        Vaginal dryness: G2 & 3.85 & 1.96 & 4.97 & 2.34 & 3.28 \\ \bottomrule
    \end{tabular}
    \caption{Percentage (\%) of selected AE term:grade classes at cycle 1 with contribution and mass at least 0.52\% by treatment in B35 trial}
     \label{tab:b35_ae_term_grade_percentage}
\end{table}



\clearpage


\bibliographystyle{bmc-mathphys} 
\bibliography{bmc_article}